\documentstyle[12pt,aaspp4]{article}

\lefthead{Bekki, Shioya, \& Couch}
\righthead{Passive spiral formation}

\begin{document}
\title{Passive spiral formation from halo gas starvation: Gradual transformation
into S0s}

\author{Kenji Bekki and Warrick J. Couch} 
\affil{
School of Physics, University of New South Wales, Sydney 2052, Australia}

\and 

\author{Yasuhiro Shioya} 
\affil{
Astronomical Institute, 
Tohoku University, Sendai, 980-8578, Japan}

\begin{abstract}

Recent spectroscopic and high resolution $HST$-imaging observations 
have revealed significant numbers of ``passive'' spiral galaxies in 
distant clusters, with all the morphological hallmarks of a spiral
galaxy (in particular, spiral arm structure), but with weak or absent
star formation. Exactly how such spiral
galaxies formed and whether they are the progenitors of present-day
S0 galaxies is unclear. Based on analytic arguments and numerical 
simulations of the hydrodynamical evolution of a spiral galaxy's halo 
gas (which is a likely candidate for 
the source of gas replenishment for star formation in 
spirals), we show that the origin of passive spirals may well be  
associated with halo gas stripping. Such stripping results mainly from 
the hydrodynamical interaction between the halo gas and the hot intracluster
gas. Our numerical simulations demonstrate that even if a spiral
orbits a cluster with a pericenter distance $\sim$  3 times 
larger than the cluster core radius, $\sim$  80 \% of the halo gas is 
stripped within a few Gyr and, accordingly, cannot be accreted by the 
spiral. Furthermore, our study demonstrates that this dramatic 
decline in the gaseous infall rate leads to a steady increase in 
the $Q$ parameter for the disk, with the spiral arm structure, 
although persisting, becoming less pronounced as the star formation rate
gradually decreases. These results suggest that passive spirals formed
in this way, gradually evolve into red cluster S0s.

\end{abstract}


\section{Introduction}

One of remarkable conclusions of observational studies
of galaxy evolution in distant (z $>$ 0.2) clusters is
that galaxies are undergoing both spectrophotometric and
morphological evolution, but the timescales appear to be
different for each (Couch et al. 1998; Dressler et al. 1999; 
Poggianti et al. 1999). In particular, these studies found
a significant number of galaxies which had the morphological
appearance of spiral galaxies, but which had no evidence of
ongoing or even recent star formation in their spectra.
The existence of these so-called ``passive'' spirals (or 
``k-type'' spirals; Dressler et al. 1999), coupled with the
strong evidence for a morphological transformation of spirals
into S0s in rich clusters since $z\sim 0.5$ (Dressler et al. 1997), 
suggests that there are two different time scales in cluster galaxy 
evolution. One is the time-scale for the suppression of star 
formation that eventually leads to spirals having passive, k-type
spectra. The other, which clearly is much longer (Poggianti et al. 
1999), is the time-scale for the morphological transformation of 
a spiral into an S0. However, it remains unclear as to (i)\,why the morphological
transformation is preceded by spectral evolution, (ii)\,what cluster-related 
physical processes are responsible for passive spiral formation,
(iii)\,how or whether passive spirals might finally be transformed into 
passive S0s, and (iv)\, whether  there is an evolutionary link between
the so-called ``E+A'' galaxies discovered by Dressler \& Gunn (1983, 1992)
and passive spirals.

In one of the first theoretical attempts to explain the 
`Butcher-Oemler' effect -- which first drew attention to these
evolutionary effects going on in rich clusters (Butcher \& Oemler 1978) --
Larson, Tinsley, \& Caldwell (1980) suggested that infall from gaseous
halos might be important for sustaining star formation in spirals,
and that these halos might be stripped in the cluster environment, 
leading to the formation of S0s. Not only might this explain the
demise of the blue, Butcher-Oemler populations in distant clusters, 
but also the presence of the smooth-armed `anemic' spirals identified
by van den Bergh (1976) in present-day clusters. More generally, 
replenishment of interstellar gas due to sporadic and continuous gas infall 
and acquisition from external environments has also been considered to 
provide reasonable and plausible explanations for the problems of 
gas consumption time-scales for the Galaxy and typical late-type galaxies 
(Kennicutt 1983), the G-dwarf problem in the Galaxy (e.g., van 
den Bergh 1962), and the formation of counter-rotating components in disk galaxies
(e.g., Bettoni, Galletta, \& Oosterloo 1991; Bertola, Buson, \& Zeilinger 1992).
Furthermore, gas accretion from extended diffuse halo gas due to radiative 
cooling is critically important for disk galaxy formation in a hierarchical 
clustering scenario (White \& Frenk 1991), although the observed diffuse X-ray 
emission from late-type spirals appears to be inconsistent with the predictions 
from this scenario (Benson et al. 2000). 

While Larson et al's halo gas stripping scenario provides a very interesting
possibility for transforming spirals into S0s, little detailed work has been done 
to understand how passive spirals fit within this framework. The purpose of this 
paper is to investigate whether the hydrodynamical interaction 
between a gaseous halo that might surround a spiral galaxy and the hot intracluster
gas is likely to lead to the formation of a passive spiral.
We do this both analytically and numerically, examining the effects the intracluster 
gas has on halo gas reservoirs of spirals orbiting a cluster. In particular, we determine how the 
total amount of halo gas stripped as a result of this interaction, depends on both 
the orbit of the galaxy within the cluster and the mass of the cluster.
We also present numerical results on the morphological evolution of spiral galaxies 
whose gas infall rates are declining as a result of halo gas stripping.
Based on these numerical results, we discuss the possible transformation
of blue spirals into red S0s via a passive spiral phase. 

\section{Halo gas stripping}

\subsection{Analytical model}

In considering the gaseous halo of a spiral galaxy embedded in a massive dark 
matter halo, we assume it is composed of diffuse hot gas (at a temperature
equal to the virial temperature of the spiral) rather than small discrete 
clouds of cold gas (e.g., high velocity clouds) as adopted in previous models 
(Bekki et al. 2001). If such a spiral falls into a cluster, its tenuous halo 
gas experiences the ram pressure ($P_{\rm ram}$) of the hot intracluster 
medium (ICM):
\begin{equation}
P_{\rm ram} = {\rho}_{\rm  ICM}  {v_{\rm rel}}^{2},
\end{equation}
where ${\rho}_{\rm  ICM}$ and $v_{\rm rel}$ are the gas density of the ICM and the 
relative velocity of the galaxy with respect to ICM, respectively. If $P_{\rm ram}$ 
exceeds the halo gas pressure, the halo gas is efficiently removed from the 
spiral's halo region. For a spiral with total mass $M_{\rm T}$ (including its dark 
matter halo), a halo gas density ${\rho}_{\rm  h}$, and a halo extension  
$R_{\rm h}$, the critical density below which the halo is efficiently
stripped is given by:
\begin{equation}
\rho_{\rm h}= 3.93\times 10^{-5}(\frac{\rho_{\rm ICM}}{5.64 \times 10^{-5} {\rm cm}^{-3}})
{(\frac{v_{\rm rel}}{1000 {\rm km s^{-1}}})}^2
(\frac{10^{12} M_{\odot}}{M_{\rm T}}) 
(\frac{R_{\rm h}}{\rm 100 kpc}) 
{\rm cm}^{-3} \;
\end{equation}
where the halo gas temperature is assumed to be proportional to $M_{\rm T}/R_{\rm h}$
(i.e., the temperature of the gas is equivalent to the virial temperature).
Recent X-ray observations of the Galaxy halo by ASCA have demonstrated that
the halo density within 100 kpc of the Galaxy is less than $7.62 \times 10^{-5}$ 
${\rm cm}^{-5}$ (Osone et al. 2000), which is very close to the above critical 
value. Therefore, the above equation implies that even if a galaxy passes through
an ICM with a density two orders of magnitude lower than the central density
of Coma ($\sim$ $5.64 \times 10^{-3} {\rm cm}^{-3}$), the galaxy halo gas can 
still be greatly affected by the ICM. This suggests that the effects of the
cluster ICM on the halo gas evolution of infalling spirals can be widespread
throughout the cluster.
 
\subsection{Numerical simulations on halo gas stripping}

In order to estimate the amount of halo gas that can be stripped from a spiral, 
numerical simulations are performed to evaluate the amount of dynamical evolution 
the gas undergoes when subjected to both the ram pressure of the ICM and
the global tidal field of the cluster. The reason for including the latter 
is that previous studies have highlighted the likely importance of this
in halo gas evolution (Larson et al. 1980; Bekki et al. 2001).  
A spiral is assumed to orbit a cluster (or a group) which has a  
total mass of $M_{\rm cl}$, a core (or scale) radius of $R_{\rm s}$,
a virial radius of $R_{\rm vir}$, and density profiles of dark halo
and hot gas equivalent to those predicted by Cold Dark Matter (CDM) models
(Navarro, Frenk, \& White 1996). The initial radial density profile (${\rho}_{\rm 
ICM}$) of the hot ICM is assumed to be the same as that observed by Vikhlinin, 
Forman, \& Jones (1999) (i.e., ${\rho}_{\rm ICM}$ $\propto$ $R^{-2.25}$).
We investigate both a {\it cluster} model with $M_{\rm cl}$ = 5 $\times$ 
$10^{14}$ $M_{\odot}$, $R_{\rm s}$ = 230\,kpc, and $R_{\rm vir}$ = 2.09\,Mpc
and a {\it  group} model with $M_{\rm grp}$ =  $10^{13}$  $M_{\odot}$,
$R_{\rm s}$ = 62.4\,kpc, and $R_{\rm vir}$ = 0.57\,Mpc. 
The orbital configuration of the spiral is specified by
($x$,$y$,$z$)=($R_{\rm in}$,0,0) and 
($V_{\rm x}$,$V_{\rm y}$,$V_{\rm z}$)=(0,$\alpha V_{\rm cir}$,0),
where $\alpha$ is a free parameter and $V_{\rm cir}$ is the circular velocity
at $R$ = $R_{\rm in}$.

We present the results of the models with $R_{\rm in}$ = $R_{\rm vir}$ and 
$\alpha$ = 0.1, 0.25, 0.5, 0.6, 0.7, 0.8, and 1.0.
The halo gas is composed of $10^5$ test particles 
and its initial spherically-symmetric radial density distribution 
($\rho (r)$) within the cut-off radius of 100\,kpc
is given as $\rho (r)$ $\propto$ $1/(r^2+{a_{h}}^2)$,
where $a_{h}$ and $r$ are the halo core radius (set to be 7\,kpc) and radius 
from the center of the spiral, respectively. Each particle is assumed to have 
an initial gas density of 3.93 $\times 10^{-5}$ cm$^{-3}$.

Each particle is subject to the gravitational forces from the {fixed} cluster and 
the spiral potentials and the ram pressure force of the ICM. 
The spiral's gravitational potential is assumed to have three components: a dark 
matter halo, a disk, and a bulge. We assume a logarithmic dark matter halo 
potential,
\begin{equation}
{\Phi}_{\rm halo}=v_{\rm halo}^2 \ln (r^2+d^2),
\end{equation}
a Miyamoto-Nagai (1975) disk,
\begin{equation}
{\Phi}_{\rm disk}=-\frac{GM_{\rm disk}}{\sqrt{R^2 +{(a+\sqrt{z^2+b^2})}^2}}
\end{equation}
and a spherical Hernquist (1990) bulge 
\begin{equation}
{\Phi}_{\rm bulge}=-\frac{GM_{\rm bulge}}{r+c},
\end{equation}
where $r$ is the distance from the center of the spiral, 
$d$ = 12\,kpc, $v_{\rm halo}$ = 131.5\,km ${\rm s}^{-1}$,
$M_{\rm disk}$ =  $10^{11}$ $M_{\odot}$,
$a$ = 6.5\,kpc, $b$ = 0.26\,kpc, $M_{\rm bulge}$ =  3.4 $\times$ $10^{10}$ 
$M_{\odot}$, and $c$ = 0.7\,kpc. This set of parameters is reasonable and 
realistic for the Galaxy. We calculate the ram pressure force of each gas 
particle at time $T$ based on equation (1) and the particle's velocity and ICM 
density at the particle's position at $T$. 

Figure 1 shows that even if a spiral is orbiting the region
well outside the cluster core ($R_{\rm p}$ $\sim$ 3.2 $R_{\rm s}$), the ram 
pressure of the ICM and the cluster tidal force can still efficiently remove the 
halo gas within a few billion years. As is shown in Figure 2, the total amount of  
stripped halo gas depends strongly on the orbital pericenter in such a way
that a larger amount of halo gas is stripped in a spiral with smaller $R_{\rm p}$.
This result suggests that if passive spirals in clusters are formed from
halo gas starvation, these passive populations are likely to be located in 
the inner cluster regions. Furthermore, Figure 2 demonstrates that the ram 
pressure effect of the ICM is much more significant than that of the cluster tidal
field in terms of halo gas stripping.

Halo gas stripping, however, is not so efficient in the group environment. The
group model, which is also plotted in Figure 2, shows that the halo gas is
rapidly stripped only if the spiral passes through the core of the group, the
only region where the density of the intragroup medium (IGM) is expected to be high.
These results imply that dynamical and hydrodynamical interaction between the 
halo gas of spirals and the intragroup medium could be  
less important in groups than in clusters.  
However, the above results imply that the star formation rates of galaxies passing through
the center of a group can be strongly affected by the IGM. 
We thus suggest that the observed difference in physical properties between field and group
galaxies (e.g., Mulchaey \& Zabludoff 1998; Tran et al. 2001), such as the relatively
suppressed star formation in {\it some} group member galaxies,
can be due partly to the halo gas stripping of galaxies.

In the above simulations, we assumed that the spiral halo gas
is composed only of diffuse hot gas; we did not  
consider that some fraction of the halo gas `lumpy' or is 
located in discrete gas clouds such 
as the high velocity clouds suggested to exist in the surrounding regions
of the Galaxy and the Local group of galaxies (e.g., Blitz et al. 1999). 
These discrete gas clouds are likely to be influenced much less 
by the hot ICM, because the ram pressure 
is less efficient for such clouds owing to their relatively compact configuration. 
Consequently, a considerably smaller amount of halo gas would be stripped if
the halo gas of spirals infalling onto a cluster is composed both of
hot diffuse gas and cold discrete gas clouds. 
We therefore suggest that the total amount of gas stripped in the present
simulations could be overestimated because of the model's not including
the possible cold gaseous clouds in the simulations.

\section{Morphological evolution of spirals after gas stripping}

In order to investigate how spiral galaxies evolve dynamically after 
halo gas stripping and the subsequent rapid decline of gas infall
onto their disks, we numerically simulate the {\it morphological} 
evolution of spirals with different infall rates.
These numerical simulations are carried out using a GRAPE board (Sugimoto et al. 
1990). In these GRAPE simulations, a spiral disk with an initial disk
mass of 3 $\times$ $10^{10}$ $M_{\odot}$ and an initial $Q$ parameter (Toomre 1964)
value of 1.0 is composed of 20,000 collisionless
particles, and (for self-consistency) these particles respond to the fixed 
gravitational force of the dark halo and bulge components described in  
equations (3) and (5). We steadily add new collisionless particles 
(which are considered to be the ``gaseous components'' here) on circular orbits
to the disk in order to mimic the accretion of dynamically cool gaseous
components from the halo region. This scheme for disk growth due to ``gas'' 
replenishment was first introduced by Sellwood \& Carlberg (1984). However, the 
present model is more realistic than theirs in the sense that it includes an 
exponential density profile and a realistic rotation curve for  the disk. 
To assess the dependence of morphological evolution on the gas accretion 
(infall) rate, 5 simulations were run each with the following different
${\dot M}_{\rm acc}$ values: 0.0, 0.9, 1.8, 4.4, and 8.8 $M_{\odot}$ yr$^{-1}$. 
The choice of these values was guided by the numerical results in the previous 
section on halo gas stripping.

In Figure 3 we show a collection of `snapshots' from these simulations, 
showing model spirals with different accretion rates at critical times in
their morphological evolution. 
In the early dynamical stages of all models,
trailing spiral arms are formed because of the ``swing amplification mechanism''
(Sellwood \& Carlberg 1984).
It can be seen that the spiral arms become 
invisible 
in the model with ${\dot M}_{\rm acc}$ = 0.0 $M_{\odot}$ yr$^{-1}$ 
within $\sim$  a few  Gyr,  
whereas in the models where the accretion rate is non-zero, the
spiral arm structure persists for longer and is more pronounced at
any given time in those with the higher rates. In the `low accretion' model with 
${\dot M}_{\rm acc}$ = 4.4 $M_{\odot}$ yr$^{-1}$, the spiral structure 
is only just visible by $T=3.4$\,Gyr, whereas it is 
much stronger in the `high' accretion model (${\dot M}_{\rm acc}$ = 8.8 
$M_{\odot}$ yr$^{-1}$) at this point. 
This trend is also followed by the other models with even
lower accretion rates (0.9 and 1.8 $M_{\odot}$ yr$^{-1}$).
These results confirm the early results by Sellwood \& Carlberg (1984)
and suggest that gas infall from halo regions, which is demonstrated by 
our simulations (in Figure 1 and 2) to be greatly affected by the ICM,  
is a key parameter for morphological evolution of spirals.

Star formation and spiral arm structure is thought to be closely
linked to the $Q$ parameter (Sellwood \& Carlberg 1984; Kennicutt 1989), 
and in Figure 4 we show how it varies with radius in the low
accretion model at $T=3.4$\,Gyr. As a result of the dynamical heating
of the spiral arms,
the $Q$ parameter steadily increases from an initial value of 1 to 1.5$-$2, in 
particular, in the inner ($r$ $<$ 5 kpc) and the outer ($r$ $>$ 12 kpc)
regions. Considering that observations indicate that the $Q$ parameter 
is less than $\sim 1.5$ for active star formation (Kennicutt 1989),
this result suggests that star formation is heavily suppressed in
the spiral with the low accretion rate. 
If each of the annular disk regions with $Q$ $<$ 1.5 are considered to be 
sites of star formation, the fraction of galaxy mass contained within 
these sites is only 4 \% at $T$ = 3.4 Gyr,
which is more than an order of magnitude smaller than that
in the early stages of the disk evolution (e.g., 75 \% at $T$ = 0.4 Gyr).

In combination therefore, Figures 3 and 4 show how a passive spiral
can develop within our halo gas stripping scenario. Taking the
`low' accretion model at $T=3.4$\,Gyr as an example, we see that it
has weak but still discernible spiral arm structure (sufficient for
it to still be classified as a spiral), but its star formation
has already dropped dramatically.
Since previous spectrophotometric studies (e.g., Shioya et al. 2002) have
demonstrated that disks that suffer an abrupt decline in star formation evolve
into passive k-type galaxies within 2 Gyr, the above result
implies that k-type spirals can be formed from spirals with low 
halo gas accretion rates. 
Moreover, our simulations also show that such a galaxy eventually evolves
into a system with a featureless disk, as seen for the `no accretion'
model at $T=3.4$\,Gyr. Such a galaxy is most likely to be classified morphologically
as an S0 with a bulge-to-disk-ratio of $\sim 0.6$.
Thus our results suggest that if the gas accretion 
(infall) rate onto a blue, gas-rich spiral gradually declines as a
result of halo gas stripping, it will eventually be transformed into a red  
S0, with the passive spiral appearance being a transitionary phase in
this process.

\placefigure{fig-1}
\placefigure{fig-2}

\placefigure{fig-3}
\placefigure{fig-4}

\section{Discussion}

\subsection{S0 formation with and without secondary starbursts}

Our study shows that the transformation in star formation rate and morphology
associated with the efficient stripping of a spiral's halo gas, involves a
phase consistent with a passive, k-type spiral. This {\it gradual} 
transformation process, during which spirals stay intact and
maintain their thin-disk components, is in striking contrast to the merger
scenario (Bekki 1998) where the coalescence of two unequal-mass disks  
{\it rapidly} transforms them into a red S0 with a thick disk and a
bigger bulge because of the triggered central starbursts and dynamical heating of
the preexisting disks. On the other hand, the `gradual' S0 formation process
proposed here maybe unable to explain the existence of a+k/k+a  (possibly poststarburst) S0s  
(Couch et al. 1994, 1998; Dressler  et al. 1999) and the color-magnitude relation of
luminous cluster S0s whose spectrophotometric properties were modelled in detail by 
Shioya et al. (2002) and Shioya, Bekki, \& Couch (2002). We can think
of these two S0 formation scenarios -- one rapid and the other gradual -- as 
being complementary in explaining the diversity observed in the structural,
kinematical, photometric, and spectroscopic properties of S0s.

An important question at this point is what observational signature is capable of
discriminating between these two different S0 formation scenarios in clusters
(e.g., Rose et al. 2001; Bartholomew et al. 2001)? Secondary starbursts in the central
regions of S0 progenitor spirals would seem a common process in the latter rapid S0
formation scenario. These would give rise to central blue colors
during the starburst and poststarburst ($<$ 1 Gyr after the onset of starburst) epochs,
and thus S0s formed in this way would have {\it positive} color gradients
(Bekki, Couch, \& Shioya 2002). Therefore, investigating the incidence of a+k/k+a S0s 
with positive color gradients gives a measure of the relative importance of the latter 
rapid S0 formation mechanism. We would also expect S0s formed by secondary
starbursts to exhibit the strongest a+k/k+a spectral signature in their
central regions. Spatially resolved integral field unit (IFU) spectroscopy of distant
cluster galaxies on 8m class telescopes will be important to verify this.
A comparison of our simulations of the spectrophotometric
and chemodynamical evolution of cluster spiral and S0s that undergo secondary starbursts 
with observational data (e.g., Bartholomew et al. 2001) will be presented in a future
paper.

Another important question is what mechanisms are responsible for starbursts in S0 progenitor
spirals, if some fraction of S0s are formed with secondary starbursts ?
Tidal interaction between galaxies (and  strong  cluster global tide)
and galaxy merging have been demonstrated to trigger  central strong starbursts
during these processes. However, recent HST morphological studies and spectroscopic
studies with   large ground-based telescopes have suggested that not all of disk galaxies  with 
(possible) intensively   star-forming  regions show central bright starburst knots
(e.g., Dressler et al. 1999).
Some of these spirals  are observed to show blobby, apparently  star-forming knots
{\it throughout} the disk, which indicates
that the above violent dynamical interaction is not always responsible for 
Sp $\rightarrow$  S0 formation.
Furthermore, such starburst candidates are not necessarily observed 
to be confined to the central regions of clusters, where dynamical interaction
are the most likely to trigger nuclear starbursts (Dressler et al. 1999). 
These observations thus imply that it is still premature to draw any conclusions
about which physical processes dominate the S0 formation with secondary starbursts.

\subsection{Similarity between high redshift passive spirals and nearby anemic spirals}

The anemic spirals that van den Bergh (1976) first discovered in the Virgo cluster, 
appear to be a class of objects that is intermediate between gas-rich normal
spiral galaxies and gas-poor S0s, suggestion that if they
finally lose all of their disk gas, they evolve to become S0s. 
van den Bergh suggested that the disk gas could be removed through   
ram pressure stripping and collisions between cluster member galaxies.  
Our numerical simulations suggest that halo gas stripping due to the Virgo clusters' 
hot ICM (Ulmer et al. 1980) is also a possible candidate which can explain the origin of
Virgo anemic spirals. However, what is not so clear is whether these nearby anemic spirals
have the same integrated spectral properties as the distant passive spirals. Nor is it
clear whether the distant k-type spirals have the same physical properties 
as the Virgo anemic spirals 
(e.g., number fraction of these populations, HI-to-luminosity-ratio, colors,
and surface HI gas density).

Regarding the above question, Duc et al (2002) found that the star formation rates estimated
from mid-infrared data for k-type passive spirals 
in Abell 1689 are  very small (at best $\sim$ 1 $M_{\odot}$ yr$^{-1}$), 
though these are higher than those
estimated from [OII] emission. Couch et al. (2001) also revealed that star formation 
of disk galaxies in distant clusters are globally and uniformly suppressed.
These two observations suggest that some distant k-type spirals have very low level star formation
which is similar to that  of some nearby anemic spirals in Virgo. 
It is however unclear whether these passive spirals are really 
similar to  nearby anemic spirals because of the lack of extensive HI observations
for distant cluster members.  
Also, it is unclear whether the spiral structures in distant passive spirals
are less remarkable compared with those in distant field ones.  
Extensive observational studies on the HI gas content in distant cluster spirals
and on the detailed spiral morphologies in these populations are thus very much welcome.

\section{Conclusions}

We have investigated, numerically, the dynamical interaction between the halo gas of
spiral galaxies and the hot ICM in order to estimate the amount of
halo gas that can be stripped from spirals which reside in rich clusters. 
We confirmed that this hydrodynamical interaction is an efficient means of 
stripping the halo gas in cluster spiral galaxies. 
We have also investigated the dynamical evolution of spiral galaxies
that have a very small gas accretion rate due to the removal of their
halo gas. 
We demonstrated that the spiral arm structure in these disk galaxies
becomes rapidly less pronounced and then eventually disappears.
Coupled with this is a dramatic drop in the star formation rate due
to the rapid increase in the $Q$ parameter -- a key quantity in determining
the occurrance of star formation in disk galaxies.
We therefore conclude that halo gas stripping caused by dynamical interaction
between halo gas and the hot ICM is a plausible mechanism for not
just S0 production in distant clusters but also the passive spirals
observed in distant clusters

This present study also demonstrated that the dramatic tidal effects 
associated with galaxy-galaxy interactions and merging and the mass distribution
of the cluster itself are not necessarily the sole cause of environmental
differences in galaxy evolution between rich clusters and the field.
Passive spirals recently discovered in distant clusters (Couch et al. 1998)
and anemic spirals and late-type disk galaxies with low levels of star
formation observed in nearby groups and clusters (van den Bergh 1976, Tran et al. 2001) 
are probably objects which provide valuable information on less dramatic but
long-term environmental effects on galaxy evolution. 
We lastly suggest that formation of some cluster S0s  can be due to this 
long-term environmental effects.

\acknowledgments

We are grateful to the referee Alan Dressler for his  comments, which
helped us to significantly improve this paper. 
K.B and W.J.C. acknowledge the financial support of the Australian Research Council
throughout the course of this work.
Y.S. thanks the Japan Society for Promotion of Science (JSPS) 
Research Fellowships for Young Scientists.

\clearpage


\begin{figure}
\figcaption{
{\it Upper left panel}: Orbital evolution of a spiral in a cluster
with $M_{\rm cl}$ = 5 $\times$ $10^{14}$ $M_{\odot}$,
$R_{\rm s}$ = 230 kpc, and $R_{\rm vir}$ = 2.09 Mpc.
The orbit during 4.5 Gyr dynamical evolution of the spiral
is given by a solid line.
The cluster core (or scale) radius is represented by a dotted circle.
The time 0, 2.3, 2.8, and 3.4 Gyr are represented by crosses
along the orbit. 
{\it Three remaining panels}:  
Time evolution of the halo gas distribution during the dynamical evolution
of the spiral.   
\label{fig-1}}
\end{figure}

\begin{figure}
\plotfiddle{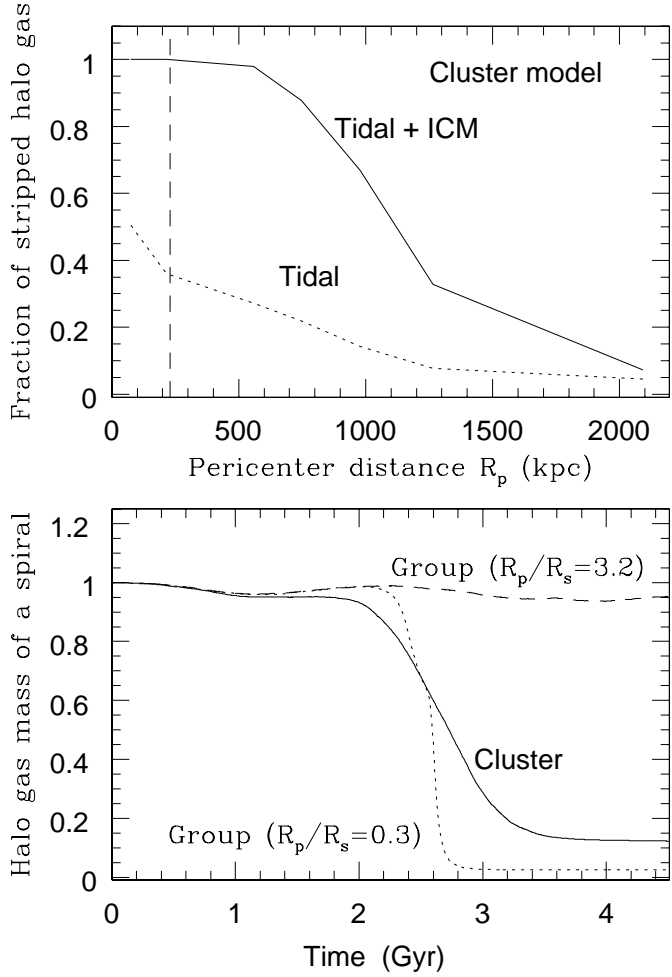}{18.cm}{0}{80.}{80.0}{-200}{0}
\figcaption{
{\it Upper}: The dependence of the fractional mass of the stripped halo gas
of a spiral on the orbital pericenter $R_{\rm p}$ for the cluster
models with (solid) and without (dotted) ram pressure effects of ICM.
The cluster model parameters are the same as those in Figure 1.
The cluster core (or scale) radius is represented by a dashed line.
Note that a spiral with smaller $R_{\rm p}$ loses a larger amount of its halo gas.
{\it Lower}: Time evolution of halo gas mass normalized to the initial mass
for a spiral orbiting the cluster  with $R_{\rm p}/R_{\rm s}$ = 3.2 (solid)
and for spirals orbiting the group  with  $R_{\rm p}/R_{\rm s}$ = 3.2 (dashed) and
$R_{\rm p}/R_{\rm s}$ = 0.3 (dotted).
\label{fig-2}}
\end{figure}

\begin{figure}
\plotone{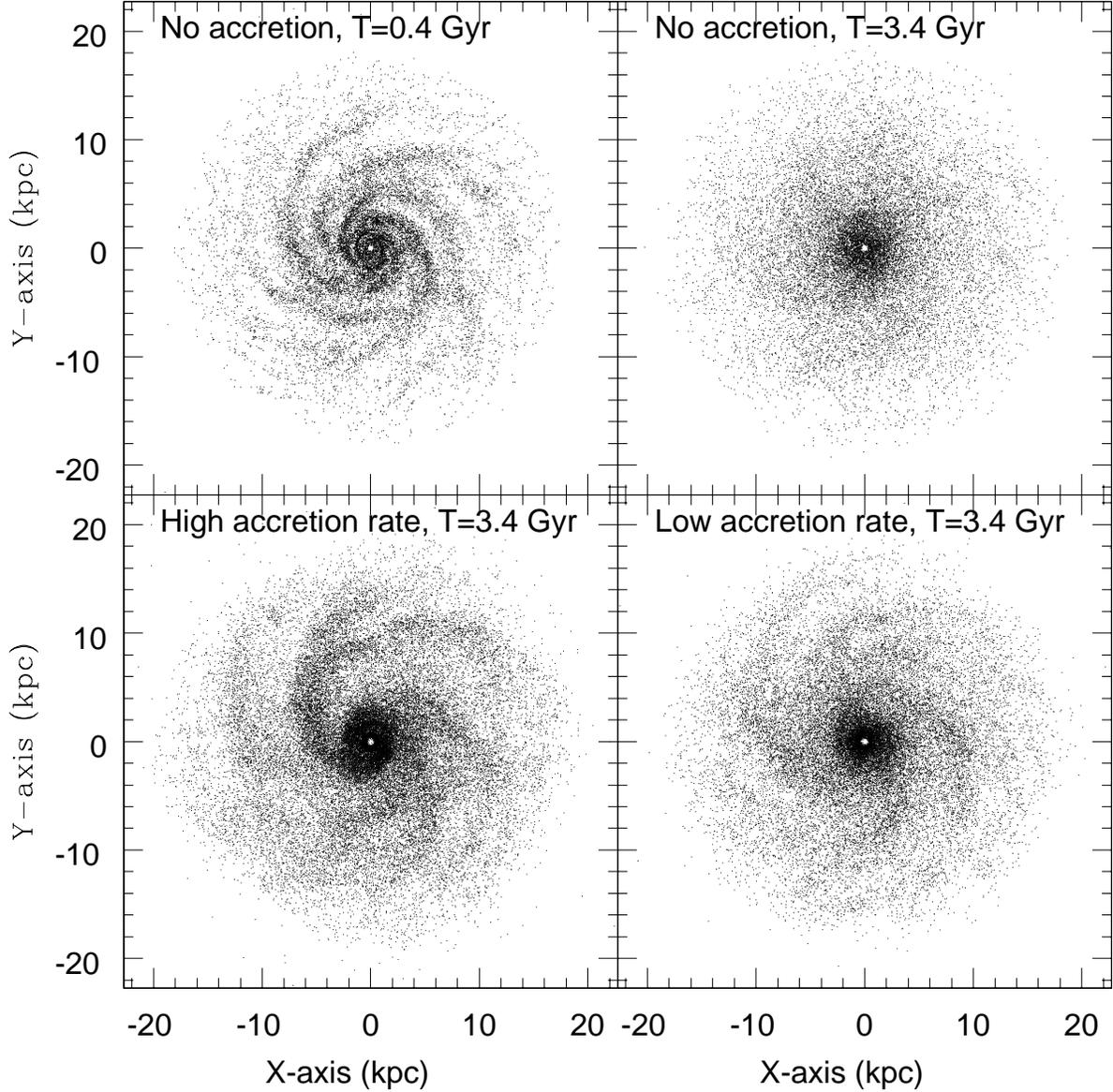}
\figcaption{
Morphological properties (face-on view) 
for models with ${\dot M}_{\rm acc}$ = 0  $M_{\odot}$ yr$^{-1}$ at $T$ = 0.4 Gyr
(no accretion, upper left),   
${\dot M}_{\rm acc}$ = 0  $M_{\odot}$ yr$^{-1}$  at $T$ = 3.4 Gyr (upper right),
${\dot M}_{\rm acc}$ = 8.8  $M_{\odot}$ yr$^{-1}$ 
at $T$ = 3.4 Gyr (high accretion rate, lower left), and
${\dot M}_{\rm acc}$ = 4.4  $M_{\odot}$ yr$^{-1}$ 
at $T$ = 3.4 Gyr (low accretion rate, lower right).
Note that spiral arms are more clearly seen in the high accretion rate model
whereas spiral arms are invisible in the no accretion
model  and less remarkable
in the low accretion rate one.
\label{fig-3}}
\end{figure}

\begin{figure}
\plotone{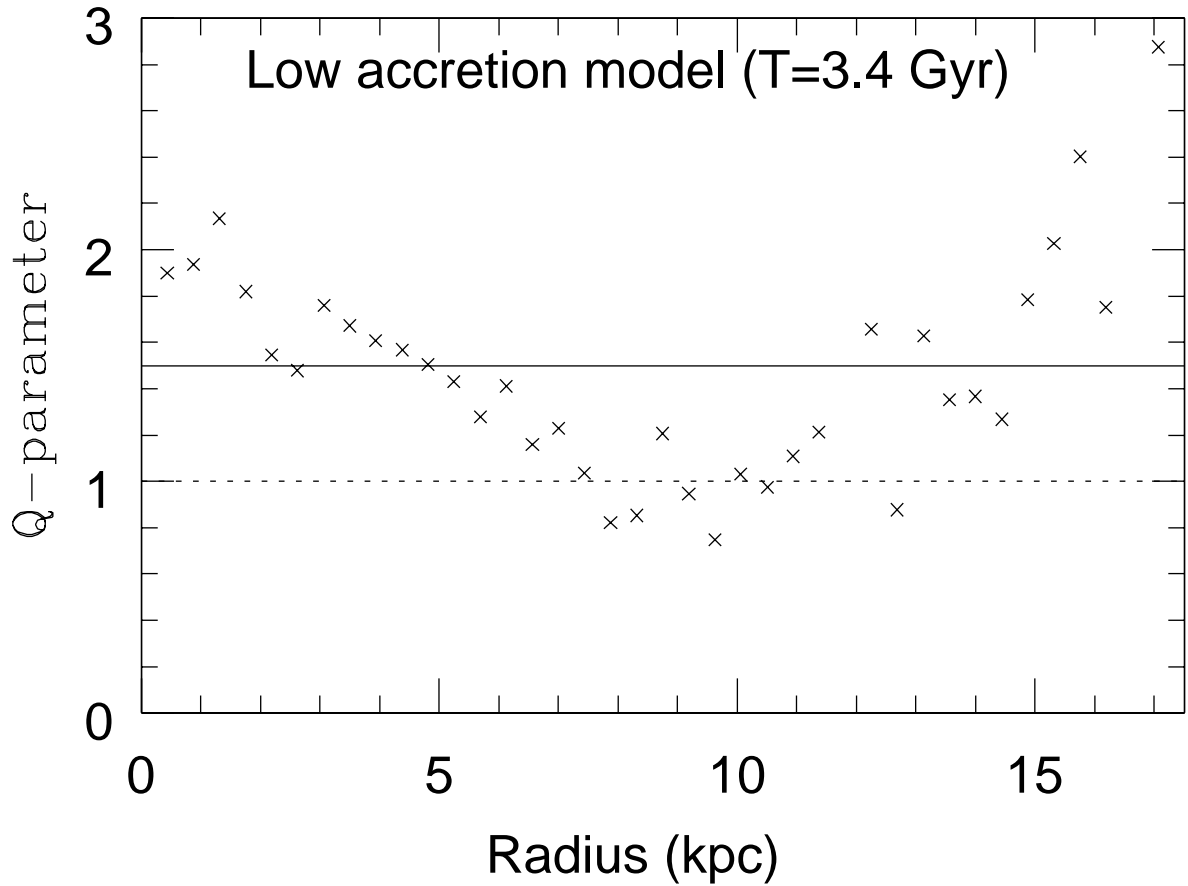}
\figcaption{
Radial distributions of the $Q$ parameter in  the low accretion rate
model with  ${\dot M}_{\rm acc}$ = 4.4  $M_{\odot}$ yr$^{-1}$ 
for  $T$ = 3.4 Gyr. Each cross represents the $Q$ value at each radial bin.
For comparison, initial value (=1) of the $Q$ parameter and the threshold one (=1.5)
for active star formation (Kennicutt 1989) are plotted by a dotted line
and by a solid one. It should be noted that only in the limited
disk regions (5$-$12 kpc),
the $Q$ parameter remains smaller 
than the threshold value ($\sim$ 1.5; Kennicutt 1989). 
This result implies that star formation can proceeds
only in the ring-like regions (5$-$12 kpc, here)
in passive spirals and S0s after the declines
of gas accretion (infall) rates. 
\label{fig-4}}
\end{figure}

\end{document}